\documentclass[12pt]{iopart}
\usepackage{graphicx}

\begin{document}

\title{Towards a modeling of the time dependence of contact area between solid bodies}


\author{E. A. Jagla}

\address{Centro At\'omico Bariloche, Comisi\'on Nacional de Energ\'{\i}a At\'omica, 
(8400) Bariloche, Argentina}

\begin{abstract}

I present a simple model of the time dependence of the contact area between solid bodies, assuming either a totally uncorrelated surface topography, or a self affine surface roughness. 
Time dependent deformation due to ``creep" processes is incorporated using a recently proposed model, and
produces the time increase of the contact area $A(t)$. For an uncorrelated surface topography, $A(t)$ is numerically found to be well fitted by expressions of the form
[$A(\infty)-A(t)]\sim (t+t_0)^{-q}$, where the exponent $q$ depends on the normal load $F_N$ as $q\sim F_N^{\beta}$, with $\beta$ close to 0.5. In particular, when the contact area is much lower than the nominal area I obtain $A(t)/A(0) \sim 1+C\ln(t/t_0+1)$, i.e., a logarithmic time increase of the contact area, in accordance with experimental observations. The logarithmic increase for low loads is also obtained analytically in this case. For the more realistic case of self affine surfaces, the results are qualitatively similar.
\end{abstract}
\maketitle

\section{Introduction}

The contact area between solid bodies is in general only a tiny fraction of the apparent (or nominal) macroscopic contact area, and is essentially proportional to the normal force between the bodies\cite{persson,bt}. This fact is at the core of many important characteristics of friction phenomena. In particular, it gives a natural explanation of the independence of the friction force with the nominal contact area, and the linear increase of friction force with normal load.
However, contact area, even under static external condition is a time dependent quantity, and this produces  time dependent effects in friction. For instance, the logarithmic time increase of static friction coefficient with the contact time, that is well described in many materials\cite{mudet,marone0}, is mostly related to a corresponding logarithmic increase in time of the contact area\cite{caroli}. Direct evidence of this geometric
aging effect was provided by Dieterich and Kilgore\cite{dk}. The 
time increase of contact area is also related with the phenomenon of velocity weakening during sliding, namely the fact that the friction force between sliding surfaces can decrease when the relative velocity is increased.
The velocity weakening effect is extremely important in relation with the dynamics of the friction process. In fact, it can be shown that this effect is at the base of the process of earthquake generation during the relative motion of tectonic plates\cite{scholz}, and also plays an important role in the stick-slip dynamics of many sliding systems\cite{persson}. 
These examples highlight the central role that the time increase of contact area has
in the description of many frictional phenomena. An understanding of the time dependence of contact area is thus
a very basic aim of any friction theory.

It is generally accepted that the physical mechanisms by which contact area increases in time are associated with
plastic 
phenomena occurring in the materials. 
For our purposes, plastic processes can roughly be classified into two qualitatively different groups\cite{plasticity,lakes}. Rapid plastic effects occur when the
imposed deformation conditions produce the overpassing of the yield stress of the material. This behavior is typically
referred to as plastic flow. For lower applied stresses, in cases in which the yield stress is not 
overpassed, there still exists the possibility of thermally activated reacommodations in the system that 
tend to reduce gradually its free energy. These processes are much slower than those of plastic flow, and are strongly temperature dependent. They are refereed to as creep phenomena. 
When two solid bodies are pressed together, the real contact occurs only in a tiny part of the surface, and the local stresses 
are correspondingly very high. This may produce local plastic flows that increase the contact area until the local stresses decay below the material yield stress. After this initial stage, the contact area typically increases slowly in time due to creep processes. The time increase of contact area in this stage is seen to be logarithmic in time.

Creep processes (that in other contexts are referred to as aging effects) produce in the contacting bodies a tendency  to reach progressively more stable configurations. The meaning of ``more stable" here, is that the atomic rearrangements implied by a creep process must produce on average a decrease in the total free energy of the system.
This stabilizing tendency originates in the fact that  energy barriers to jump onto a lower energy state are in general lower than those to go to higher energy states, so on average creep produces always an energy reduction in the system. Along these lines, the phenomenon of contact area increase has been qualitatively described in some idealized geometries\cite{brechet,caroli}, and the experimental logarithmic increase law has been justified for these cases. Crucial in this analysis is the assumption of a phenomenological creep law in which some generic strain rate $\dot s$ is exponentially related to an appropriate stress $\sigma$, i.e., $\dot s\sim \exp(\sigma/S)$, where $S$ is called the strain rate sensitivity and is temperature dependent \cite{caroli}.
The modeling of the phenomenon in more realistic cases is prevented by the complication
to describe creep in the materials in a sensible and analytically (or even numerically) tractable way.
As far as I know there is no statistical model that, based on well defined microscopic evolution laws, is able to predict
the logarithmic increase of contact area with time. 

Recently, a way to incorporate creep effects in the dynamics of sliding friction has been proposed in \cite{uno,dos}. In particular this has been applied to models (the Burridge-Knopoff\cite{bk} and Olami-Feder-Christensen\cite{ofc} models) used to describe seismic phenomena.
This kind of models describe the friction phenomena occurring at the flat (on average) contact surface between two solid bodies that are sheared against each other. 
The kind of modeling proposed in \cite{uno,dos} consists in defining some ``plastic" degrees of freedom, and arguing that they evolve with a tendency to minimize the total energy of the system, which is a generic realization of creep processes as  described in the previous paragraph.
It was shown that the proposed relaxation mechanism is able to generate realistic sequences of earthquakes, a goal that had not been obtained previously without this kind of modification. Also, realistic frictional properties
are well reproduced using this relaxation mechanism. In particular, a logarithmic increase of the static friction coefficient with contact time, and an approximately logarithmic decrease of the average friction force as a function of relative velocity has been obtained.


In view of the standard interpretation of macroscopic friction features in terms of contact area, and since the structural relaxation mechanism in \cite{uno,dos} is consistent with macroscopic friction properties, the question arises if there is
a way to use that relaxation mechanism to model the time increase of contact area of solid bodies under static contact. Such a modeling, if successful, would give further support to the structural relaxation mechanism, and would provide an appropriate framework to study the phenomenon of time increase of contact area in greater detail than the rather qualitative descriptions available up to now.
I attempt this particular goal here.
In order to do this I explain a variation of the models presented in \cite{uno,dos} that allows to define the contact area, and at the same time incorporates the structural relaxation mechanism. I show that the main phenomenology associated to the contact area is re-obtained, particularly, its logarithmic time increase.

\section{The Model}

Typically, the surfaces of solid bodies in contact are at the same time rough and elastic (or elasto-plastic), and this is important when studying sliding friction. In the context of static contact however, some simplifying assumption can be made. I will consider the case of an elastic surface that is perfectly flat down to atomic scale in the absence of external forces, and an opposing surface that is atomically rough, but strictly rigid.
The two surfaces are oriented horizontally. The underlying rigid rough surface is described by a random variable defined over the plane $\xi({\bf r})$. In the numerical simulations the values of ${\bf r}$ will be restricted to lie on a two dimensional square mesh.
Two cases will be considered separately: one in which the $\xi({\bf r})$ are drawn from a unique Gaussian distribution
independently for each value of ${\bf r}$, and a second case in which the $\xi({\bf r})$ is spatially correlated in order to model a self-affine surface (see below). 

A realistic modeling of the upper elastic surface should consist in principle in determining the equilibrium values of a three dimensional vector displacement field ${\bf u}$ depending upon the two horizontal coordinates, under the action of the surface forces, taking into account the elastic response of the surface, measured by an appropriate response function. This is what a full contact mechanics calculation aims to. In the present case, and already foreseeing the further inclusion of creep effect, some drastic simplifications have to be made.

The simplified description of the elastic surface (sketched in Fig. 1)
consists of a collection of scalar coordinates $u({\bf r})$, representing the vertical positions of the elastic surface at point ${\bf r}$, which are coupled via elastic springs connecting nearest neighbor mesh points, and elastic interactions with the bulk of the material, defined by a set of coordinates $u^0({\bf r})$. Note that the only degrees of freedom I allow are vertical displacements, and along this direction all the springs act. 
This modeling of the elastic surface is highly simplistic, but at this point this is necessary in order to have a solvable model. Note in particular that the consideration of only vertical displacements means that we are dealing with a material with ``zero Poisson ratio".
Also, it can be seen that a localized force onto this surface generates a distortion that decays exponentially with distance, whereas the true response of a semi-infinite elastic body is known to decay as $\sim r^{-1}$.
The limitations and some unrealistic features of the elastic model I am using are described in more detail in the last section of the paper.

The restriction imposed by the contact geometry is that $u({\bf r})\geq \xi({\bf r})$.
Given a set of values $u^0({\bf r})$, the elastic energy of the system
is 
\begin{equation}
E=\frac{k_0}{2}\sum_{({\bf r},{\bf r}')} [u({\bf r})-u({\bf r}')]^2 +\frac{k_1}{2}\sum_{{\bf r}}[u^0({\bf r})-u({\bf r})]^2
\label{e}
\end{equation}
where $({\bf r},{\bf r}')$ stands for pairs of neighbor sites on the numerical mesh. In all cases periodic boundary conditions will be used.
The equilibrium values of $u$'s are thus found by solving the set of equations obtained by minimizing Equation (\ref{e}), namely
\begin{equation}
k_0(\nabla^2 u)({\bf r})+k_1[u^0({\bf r})-u({\bf r})]=0
\label{us}
\end{equation}
(where $\nabla^2$ is the discrete Laplacian operator on the square lattice, lattice parameter is taken as unit of length) with
the constraint imposed by the contact condition $u({\bf r})\geq \xi({\bf r})$.
The number of points for which $u({\bf r})= \xi({\bf r})$ is the number of contact points, and I will take this number as a measure of the contact area $A$ in the model.
The global vertical position of the elastic surface can be measured by the mean value $\overline {u^0}$ of the $u^0$'s. 
Note also that $\overline {u^0}$ can be interpreted as a measure of the nominal distance between the two bodies and that the normal force $F_N$ is the sum of the forces of all vertical springs, i.e, $F_N=k_1\sum_{\bf r} (u({\bf r})-u^0({\bf r}))$. 

Conceptually in the same manner as in previous work\cite{uno,dos}, plastic relaxation is incorporated through a time dependence of the values of $u^0({\bf r})$. The form of the evolution will be obtained from the prescription that the energy of the system $E$ tends to be reduced during relaxation. Concretely, I use
\begin{equation}
\frac{du^0({\bf r})}{dt}=R\nabla^2\frac{\delta E}{\delta u^0({\bf r})}=k_1 R\nabla^2 (u^0({\bf r})-u({\bf r}))
\label{u0s}
\end{equation}
i.e, a standard relaxation equation that tends to reduce the value of $E$ over time as much as possible. The time scale for relaxation (controlled by $R^{-1}$) will be assumed to be much larger than the elastic time scale in which the elastic variables $u$ accommodate to satisfy Eq. (\ref{us}).
Note that the meaning of Eq. (\ref{u0s}) is that relaxation makes the force exerted by the vertical springs tend to an uniform value when $t\to\infty$. 

\begin{figure}
\centerline{\includegraphics[width=.5\textwidth]{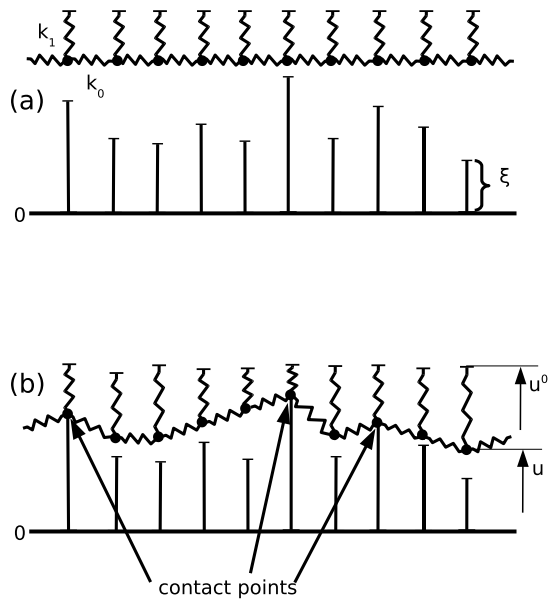}}
\caption{One dimensional sketch of the model. (a) A flat and elastic surface (defined by the black dots) on top of a rigid rough one (defined by the top of the vertical segments), before contact.(b) The situation after contact.
Dots [with vertical coordinates $u({\bf r})$] are in equilibrium under the action of vertical and horizontal springs, and eventually the force exerted by the rigid rough surface defined by $\xi({\bf r})$. The coordinates $u^0({\bf r})$ evolve in time according to the relaxation equation (\ref{u0s}). The external control parameter is the mean value $\overline{u^0}$.
}
\label{f1}
\end{figure}

The variable that is assumed we can control is $\overline {u^0}$. In this respect, note that this variable is not modified by the time evolution [Eq. (\ref{u0s})]. Actually, this is one of the reasons to use a conserving dynamics, in which the Laplacian operator is introduced in Eq. (\ref{u0s}), instead of a non-conserving one, in which the Laplacian is absent\cite{chl} (in connection with this choice, see also the final Section of the paper).
To describe a possible experimental situation, I first assume (Fig. \ref{f1}(a)) that $\overline {u^0}$ is very large, in such a way that there is no contact between the two surfaces at any point. Allowing infinite time to relax under this condition, the system reaches the uniform state in which $u^0({\bf r})=\overline{u^0}$ and $u({\bf r})=\overline{u^0}$ everywhere. This is in fact the most relaxed configuration since the elastic energy of every spring is zero. I then place $\overline {u^0}$ at some value in which contact occurs at some positions (Fig. \ref{f1}(b)), and solve Equations (\ref{us}) and (\ref{u0s}) as a function of time.
Numerically, the procedure consists in advancing the solution of Eq. (\ref{u0s}) for $u^0$ by one time step, solve Eq. (\ref{us}) using a standard relaxation algorithm\cite{numrec} for the new values of $u$, and iterate the process.

\section{Results for uncorrelated roughness}

I first show results in the case in which the variables $\xi({\bf r})$ that describe the roughness of the underlying surface are taken independently at each site, from a Gaussian distribution of zero mean and unitary variance. Although this case is not very realistic, we will see that in addition to the possibility of accurate numerical simulation, it allows for very insightful analytical treatment.
Results corresponding to $R=0$ are presented in Fig. 2, where the contact area $A$ (i.e., the number of points in contact) is plotted against $F_N$. Curves for different values of the ratio $k_1/k_0$ are shown.
Two main regimes are observed. For low normal load the contact area is essentially proportional to the load, whereas if load is too high, we reach a regime of full contact. 
The crossover between partial contact and full contact occurs at a value of $F_N$ that depends on the elastic constants of the model. If one of the spring constant dominates over the other the crossover value of $F_N$ is proportional to the dominating spring constant. For instance, if $k_0$ is negligible compared to $k_1$, the elastic surface becomes a collection of independent springs (in other contexts this kind of description of an elastic surface is described as the Winkler model\cite{winkler}), and the crossover to full contact occurs for a normal load per spring of order $k_1\sigma$, where $\sigma$ is the typical roughness of the surface, that is taken as 1 in the present simulations.

Given the equilibrium configuration for some value of $F_N$, we can set a finite value of $R$ and follow the evolution of the contact area in time. In this process it has to be taken into account that if we keep $\overline {u^0}$ fixed, the value of $F_N$ will change in time. Since experiments are usually done at constant normal force instead of constant relative distance, I implemented a feedback loop in the simulation that allows to keep the value of $F_N$ as constant by changing $\overline{u^0}$.  

The asymptotic ($t\to \infty$) value of the contact area is determined (according to Eq. \ref{u0s}) by the condition that forces on all springs $k_1$ are equal\cite{notita}. In this situation the actual value of $k_1$ is irrelevant, and the fully relaxed contact area becomes a function of $F_N/k_0$ (Fig. 1, open symbols).
Note that this asymptotic value of the contact area is not in general equal to the nominal area, i.e., the system does not reach full contact unless $F_N$ overpasses a crossover value that is proportional to $k_0$. In particular this means that if we consider the simplest case $k_0=0$ we obtain the unrealistic result that the contact area tends to the 
nominal value as $t\to\infty$. This is the reason that makes mandatory the use of a lateral spring $k_0$ in the model.
In view of standard experimental conditions, below I will concentrate in cases in which $F_N$ is such that the contact area is only a small fraction of the nominal one.
I found that the dependence of the asymptotic contact area with normal force at low loads follows a power law  $A(t\to \infty)\sim F_N^p$, with an exponent $p\sim 0.8$.

\begin{figure}
\centerline{\includegraphics[width=.5\textwidth]{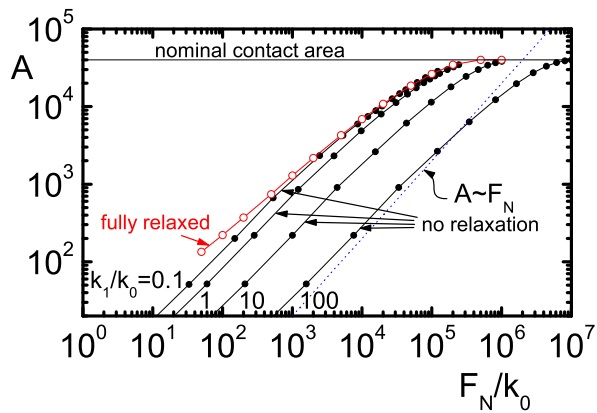}}
\caption{
Contact area as a function of normal load, for different values of $k_1/k_0$ without relaxation (full symbols), and the fully relaxed state (open symbols), in which all the forces on the $k_1$ springs are equal, in a system of 200 $\times$ 200 elements. Note that the relaxed state has always a contact area larger than unrelaxed states with the same normal force. For reference, a dotted line with a slope of 1 is also plotted.}
\label{f2}
\end{figure}

\begin{figure}
\centerline{\includegraphics[width=.5\textwidth]{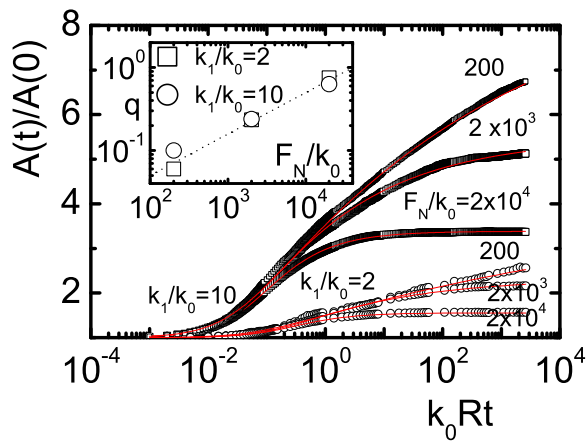}}
\caption{Time evolution of the contact area, for different values of $F_N$, and two values of the ratio $k_1/k_0$, for a system of 200$\times$200 sites. The corresponding fitting with the expression given in the text (Eq. \ref{a}) is shown by thin continuous lines. The exponent $q$ in the fitting function is plotted as a function of $F_N$ in the inset, where it is seen that $q$ goes to zero with $F_N$ as a power law (dotted line in the inset has a slope of 0.5).
}
\label{f3}
\end{figure}

The curves in Fig. \ref{f2} correspond either to ``instantaneous" or to ``fully relaxed" values of the contact area.
As a function of the contact time, the values of the contact area must evolve between these two limits.
First of all, note from Fig. \ref{f2} that the fully relaxed contact area is always larger than the unrelaxed configuration with the same normal force. This means that contact area will increase in time when relaxation is acting, which is the expected result.
The full temporal evolution of the contact area obtained by solving Eqs. (\ref{us}) and (\ref{u0s}) is shown in Fig. 3 for different values of the normal force $F_N$. I have found that all curves of Fig. 3 are very well described by expressions of the form 
\begin{equation}
A(t)/A(0)=a-\frac{a-1}{(t/t_0+1)^q},
\label{a}
\end{equation}
i.e., they indicate a saturation towards the asymptotic value with the form of a power law.
The values of $a$, $q$, and $t_0$ depend on $F_N$. In particular, the dependence of the exponent $q$ on $F_N$ is seen in an inset in Fig. \ref{f3}. It is remarkable that $q$ has a dependence on $F_N$ of the form $q\sim F_N^{\beta}$ with $\beta$ being close to $0.5$. This means in particular that for $F_N\to 0$, we can approximate Eq. (\ref{a}) by
\begin{equation}
A(t)/A(0)=1+C\ln(t/t_0+1) ~~~~~{\mbox {for}}~~ F_N\to 0
\label{log}
\end{equation}
with $C=(a-1)q$. It is numerically found that the value of $C$ is not singular in this limit.
Since experimentally the values of $F_N$ are typically tiny compared to those necessary to reach full contact, we can say that
Eq. (\ref{log}) shows that in general we must expect a logarithmic increase of the contact area in time. 

\section{Results for Self-Affine Surfaces}

The results of the previous section were obtained using uncorrelated asperities. Numerically, there is no additional complication in trying more realistic distribution of surface roughness (although this will preclude analytic treatments as that of the next Section).
Real surfaces are in fact better described as self affine fractals\cite{fract}, and are characterized in terms of its Hurst exponent $H$. This exponent measures the decaying in wave vector of the spectral distribution of surface roughness. In this section I present numerical results using a self affine surface, and show that the results are qualitatively similar to those of the previous section.

\begin{figure}
\centerline{\includegraphics[width=.5\textwidth]{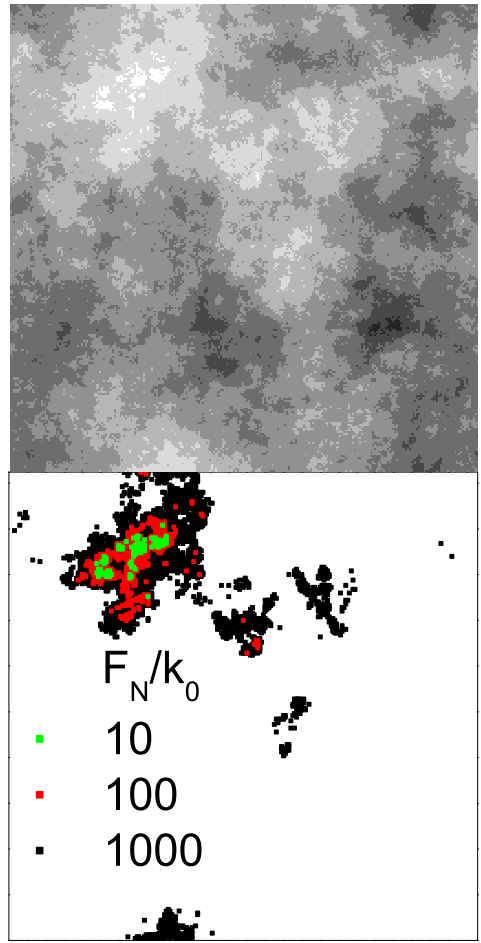}}
\caption{(a) Topography of one realization of a self affine surface ($H=1/2$) in a a system of size 256 $\times$ 256 sites and $k_1/k_0=10$. Surface height goes form -0.3 to 0.3 from back to white. (b) Contact surface as a function of load, in the absence of relaxation.
}
\label{f51}
\end{figure}

\begin{figure}
\centerline{\includegraphics[width=.5\textwidth]{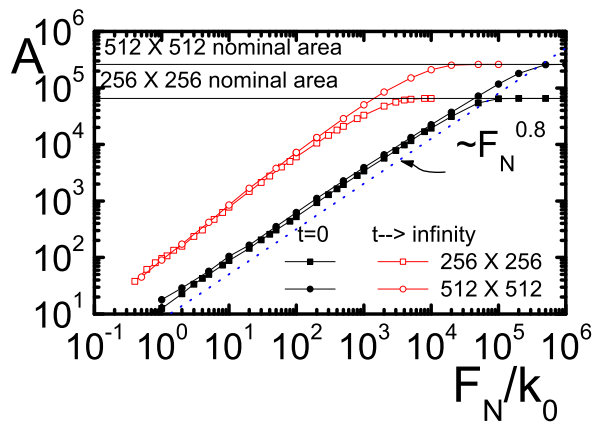}}
\caption{Contact area for a self affine surface with Hurst exponent $H=1/2$, using $k_1/k_0=10$. Solid symbols: Contact area as a function of normal load without relaxation, for two system sizes. A power law with exponent 0.8 is plotted as a reference. Note the independence of the contact area on system size in the region away from full contact.
Open symbols: corresponding fully relaxed states for the two system sizes analyzed.}
\label{f4}
\end{figure}

I construct the self affine rough surface using the successive random mean point algorithm of Voss \cite{voss,robbins}. 
The surface is characterized by its Hurst exponent $H$ and its small scale rms roughness $\Delta$.
The algorithm for the definition of the self-affine rough surface proceeds as follows\cite{robbins}: given a mesh of size $L\times L$ (for convenience $L$ is chosen to be a power of 2) the central point of the mesh is given a value of $\xi$ chosen at random from a Gaussian distribution with zero mean and width ${\ell}^H\Delta$, where ${\ell}=L/\sqrt{2}$ is the distance from the center to the corners. This center now becomes a corner of four new squares rotated by 45 $^\circ$ and with a new center-to-corner distance ${\ell}$ smaller by a factor $\sqrt{2}$. The value of $\xi$ at the center of each new square is obtained as the average of $\xi$ at the four corners plus a random value chosen from a Gaussian of width ${\ell}^H\Delta$. The process is iterated down to ${\ell}=1$. This algorithm produces a surface that is self-affine in the spatial scale between the mesh size and the full system size.
A Hurst exponent $H=0.5$ will be used (tests using other values of $H$ show no qualitative differences). I also use the value $\Delta=0.01$. 
An example of the kind of surface obtained by this method is presented in Fig. \ref{f51}(a).
Once the underlying rough surface $\xi({\bf r})$ is defined in this way, the contact with the elastic surface is numerically evaluated exactly by the same methods used in the previous Section.
The actual contact area in the absence of relaxation and for different values of $F_N$ can be seen in \ref{f51}(b).
This looks qualitatively similar to the results obtained using more realistic elastic surfaces, as that made in Ref.
\cite{robbins}.
In Fig. \ref{f4} I present results for the dependence of the contact area on $F_N$ in the present model.
In the absence of relaxation, the dependence can be fitted by an expression of the form $A \sim F_N^q$, with $q\simeq 0.8$, i.e, is a slightly sub-linear dependence. Other numerical analysis of this problem, like those in Ref. \cite{robbins}
have obtained a linear dependence.
I attribute this slight discrepancy to the somewhat artificial description of the elastic surface in the present approach, compared with the more precise, truly three dimensional description in Ref. \cite{robbins}.

\begin{figure}
\centerline{\includegraphics[width=.5\textwidth]{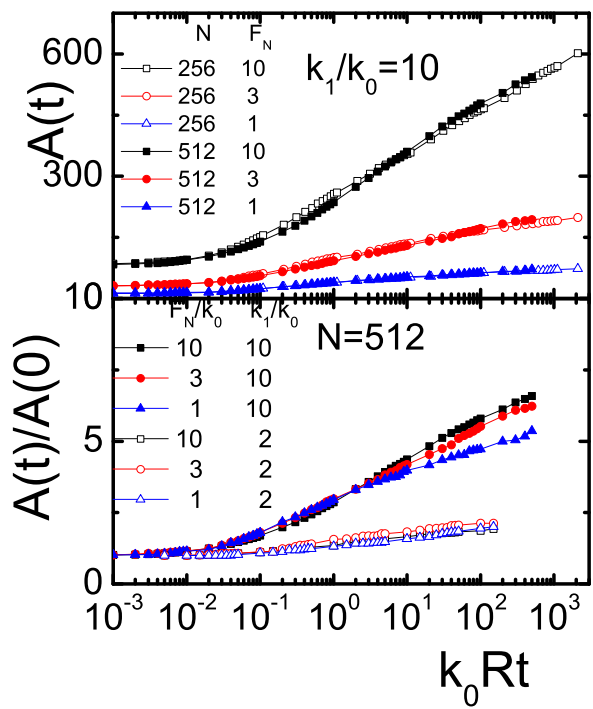}}
\caption{Time evolution of the contact area in the presence of relaxation for a self affine surface with $H=1/2$ (a) Un-normalized values for different normal forces and system sizes. (b) Results in units of the initial area for two different values of $k_1/k_0$. The results in (b) are well approximated by a logarithmic time increase with a slope that is independent of normal force, and roughly proportional to the value of $k_1/k_0$.
}
\label{f6}
\end{figure}

Now I will consider the effect of relaxation. 
The value of the limit contact area, i.e, the contact area at infinite time as a function of the normal force can be seen also in Fig. \ref{f4}. This curve, as in the uncorrelated case, was
obtained in a simulation in which a constant force on each of the vertical springs is imposed. 
The results for the increase of contact area with time are presented in Fig. \ref{f6} (note that the values of $F_N/k_0$ correspond, according to Fig. \ref{f4} to cases in which the contact area is a small fraction of the nominal one). 
We see that qualitatively the behavior is very similar to the previous case. 
In particular, for the case of very light loads and except for very long times, a very good fitting is provided by an expression as that given by Eq. (\ref{log}), with $C$ being independent of the external load.

\begin{figure}
\centerline{\includegraphics[width=.5\textwidth]{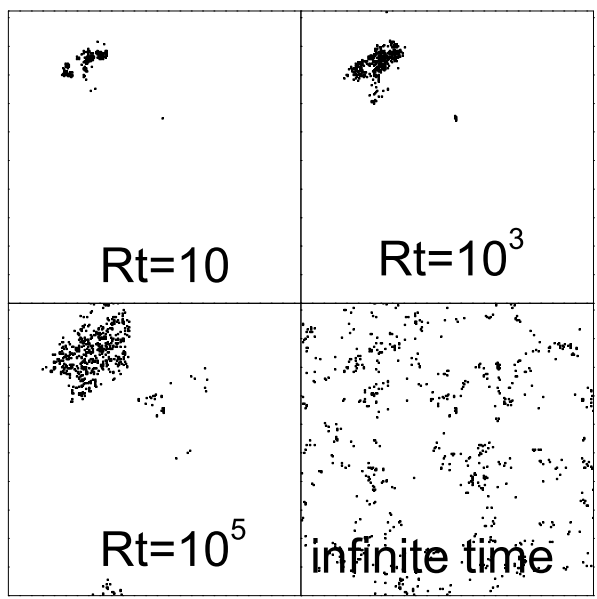}}
\caption{Time evolution of the contact surface for $F_N/k_0=10$, $k_1/k_0=10$. System size is 256 $\times$ 256. The final configuration in the last panel ($t\to \infty$)
is obtained from an independent simulation, as explained in the text.
}
\label{f52}
\end{figure}

We can see the actual contact surface 
at different times in Fig. \ref{f52}. We observe how the increase of contact area involves both an increase of the area of individual contact spots, and the appearance of new ones. This trend is very similar to the experimental findings of Dieterich and Kilgore\cite{dk}. The asymptotic surface of contact (which cannot be typically accessed experimentally), shown in the last plot in Fig. \ref{f52}, displays a uniformly scattered distribution of contact points. In particular, points that were at contact in the first stages of the process may become detached at very long times, due to the stress redistribution that occurs due to relaxation.

\section{Analytical Results}

In the limit of very light loads, the contact between surface and substrate occurs only in very few points. If 
we are considering the case of uncorrelated surface roughness, these contact point  will typically be well separated spatially.
This allows for an analytical solution of the model in this limit. We will confirm in this way the logarithmic increase of contact area in time. The analytical treatment I present turns out to be formally similar to that made in the
viscoelastic Greenwood-Williamson model \cite{caroli,hui,gauss-exp,ronsin}. In fact, the creep phenomena I am modeling through the relaxation mechanism is a kind of viscoelastic relaxation \cite{lakes,fischer-c1,fischer-c2}.
I will come back to this point by the end of the paper.

The strategy to find the solution in this limiting case is to calculate the response of the system in an ``indentation hardness test"\cite{fischer-c2}, and then exploit the linearity of Eqs. (\ref{us}-\ref{u0s}) to find the full solution.
In fact, due to their linearity, Eqs. (\ref{us}-\ref{u0s}) can be solved by Fourier decomposition.
In the case there is a single contact point between the two surfaces (supposed to be the coordinate origin) on which the force is zero for $t<0$, and takes some constant value $f_0$ for $t>0$, a direct calculation of the Fourier modes $\tilde u_q$ gives:

\begin{equation}
\tilde u_q=f_0 \left [ \frac{1}{k_0q^2}+\left (\frac{1}{k_1+k_0q^2}-\frac{1}{k_0q^2} \right)\exp[-z_qt]\right]
\label{uq}
\end{equation}
where
\begin{equation}
z_q=\frac{Rq^4k_0k_1}{k_1+q^2k_0},
\end{equation}
and the initial condition $u^0({\bf r})=0$ at $t=0$ (corresponding to a totally relaxed configuration {\em before}
contact) has been used.

\begin{figure}
\centerline{\includegraphics[width=.5\textwidth]{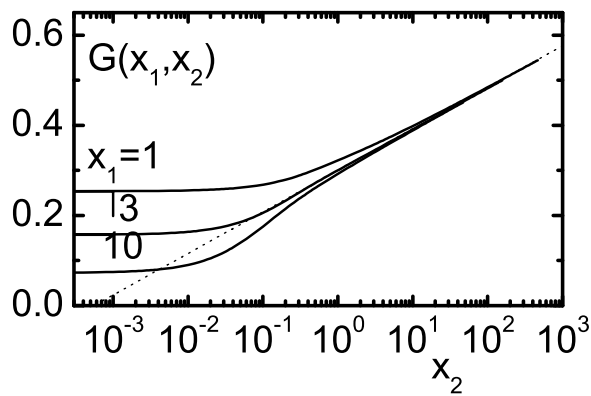}}
\caption{The dimensionless $G(x_1,x_2)$ function (Eq. (\ref{u})) as a function of $x_2$, for different values of $x_1$. Note the logarithmic increase for large $x_2$, independent of the value of $x_1$. Dotted line is the function $a_0 \ln (x_2)+a_1$, with $a_0=0.04$, $a_1=0.3$.
}
\label{funciong}
\end{figure}

By Fourier inverting this expression we can obtain the time evolution of the surface under this indentation condition. Of particular importance will be the time evolution of the variable $u$ at the contact point, i.e, $u({\bf r}=0,t)$, and its velocity $v_0(t)\equiv du({\bf r}=0,t)/dt$.
The structure of Eq. (\ref{uq}) allows to write the solution in the form

\begin{equation}
u({\bf r}=0,t)=\frac{f_0}{k_0}G\left(\frac{k_1}{k_0},k_0Rt\right)
\label{u}
\end{equation}
and
\begin{equation}
v_0(t)=f_0RG'\left(\frac{k_1}{k_0},k_0Rt\right)
\label{v0}
\end{equation}
where $G(x_1,x_2)$ is a dimensionless function of two variables, and $G'(x_1,x_2)\equiv dG/dx_2$. In Fig. \ref{funciong} I show the values
of $G$ as a function of $x_2$, for a few different values of $x_1$. We see that for $x_2$ sufficiently large (this means according to Eq. (\ref{u}), for $t$ sufficiently large) the function has a logarithmic increase, with a slope that is independent of other parameters of the model.
The form of the function $G$ is all we need to get a full solution of the contact problem on the assumption of few well separated contacts.

In the contact geometry, the value of $u({\bf r}=0,t)$ must be constant and equal to $\xi({\bf r}=0)$. In order to maintain the contact force equal to $f_0$, we have to adapt $\overline {u^0}$ accordingly,  namely
we must have
\begin{equation}
\frac{d\overline {u^0}}{dt}=-v_0(t)
\label{du0}
\end{equation}
To solve the problem of many contact points (in the assumption  that the distance among them is large compared to the size of the distortion that any contact exerts onto the surface)
we first need to generalize the form of $v_0(t)$ for a constant force $f_0$, to a new function $v(t)$ for an arbitrary time dependent form of the contact force $f(t)$. This can be done from the previous solution because of the linearity of the equations. The result is

\begin{equation}
v(t)=\int_{-\infty}^{t} d\tau R\frac{df(\tau)}{d\tau}  G'\left(\frac{k_1}{k_0}, k_0R(t-\tau)\right)
\label{vt}
\end{equation}
To generalize Eq. (\ref{du0}) to the case of many contacts, under the action of a total normal force $F_N$,
I use the fact that this normal force has to be distributed among all contacts at any time.
We obtain

\begin{equation}
\frac{d\overline {u^0}}{dt}=\frac{\sum_{m} v_m(t)}{N(t)}
\end{equation}
where $N(t)$ is the number of points at contact at time $t$, $m$ labels the contact points, and $v_m(t)$ is expression (\ref{vt}) calculated with the particular (still unknown) form of $f_m(t)$ for the contact force at point $m$. Using Eq. (\ref{vt}) and the fact that all forces must sum up to $F_N$, we obtain the simple result
\begin{equation}
\frac{d\overline {u^0}}{dt}= \frac{v_0(t)}{N(t)}
\end{equation}
where $v_0(t)$ is given in Eq. (\ref{v0}).
Namely, the change of the control variable is equal to a prescribed function of time, divided by the actual number of contacts at that time.
From here, and assuming a generic distribution of asperities given by a probability distribution $P(\xi)$, the following equation for the time evolution of the number of contacts can be derived:

\begin{equation}
\frac{dN(t)}{dt}=N_0P(\overline {u^0}(t))\frac{d\overline {u^0}}{dt}=\frac{N_0P(\overline {u^0}(t))v_0(t)}{N(t)}
\end{equation}
where $N_0$ is the total number of mesh points in the system. Integrating, one obtains
\begin{equation}
\int_{N(t=0)}^{N(t)} \frac{N(t')}{P(\overline {u^0}(t'))}dN(t')=N_0\int_0^t v_0(t')dt'
\end{equation}
The integral on the left of this equation is explicit, but it is not analytic if the $P$ distribution is taken to be a Gaussian. We can replace the Gaussian form of $P$ by an exponential to obtain a closed form, i.e., assuming $P(\xi)=\gamma \exp(-\gamma \xi)$, we obtain
\begin{equation}
N(t)=N(t=0)\left (1+k_1\int_0^t v_0(t')dt' \right).
\label{ndet}
\end{equation}
where $N(t=0)$ is the number of contacts at $t=0$, and is given (for exponentially distributed asperities) by
$N(t=0)=\gamma f_0/k_1$. 
Using the asymptotic form of $v_0$ (Eqs. (\ref{u}),(\ref{v0}), and Fig. \ref{funciong}), we obtain for large $t$
\begin{equation}
\frac{N(t)}{N(t=0)}\simeq\frac{a_0 k_1}{k_0}\ln \left(k_0Rt\right).
\label{ndet2}
\end{equation}
with $a_0\simeq 0.04$.
As numerically shown by Greenwood and Williamson\cite{gauss-exp} the consideration of a Gaussian distribution of asperities does not alter this result in the load range in which $1\ll N(t)\ll N_0$.
In this way, we obtain analytically a result that is nicely compatible with the numerical results: for small loads, the contact area increases logarithmically in time, and once scaled with the value of the area at $t=0$, the result is independent of the precise value of the applied force.
Note also that the numerical results (Figs. \ref{f3} and \ref{f6}) show an increase of the logarithm prefactor when $k_1/k_0$ is increased, compatible with Eq. (\ref{ndet2}).
The main 
hypothesis to derive Eq. (\ref{ndet2}) have been the uncorrelated distribution of asperities, and the fact that elastic distortions of the surface at the contact points do not influence other contact points, and this means that the contact points have to be sufficiently away from each other. I will analyze this expression further in the last section of the paper.

\section{Summary and discussion}

In this paper, a simplified model for the time evolution of the contact area between solid bodies in contact has been presented. The goals of the model are twofold. On one side, for very light loads and in the case of a totally uncorrelated surface roughness, the model can be worked out analytically, and it can be shown that a logarithmic dependence of the contact area with time emerges. This means that we are able to go all the way from a well defined statistical model to its solution, and obtain logarithmic aging. Secondly, numerical simulations can be done in cases where the applied load is not necessarily small, and systematic dependences of the contact area in time others than logarithmic have been found in this case. Also, the model allows to study more realistic cases in which the roughness of the surface is assumed to be correlated.
I now focus on a discussion of the process of logarithmic increase in time of contact area, that is the most directly comparable with available experimental results.

The analytical results of the previous Section provide the clearest understanding of the origin of such a logarithmic increase within the framework of the present model. In fact, this logarithmic increase is originated in the form of the surface response to a localized and constant applied load. This can be rephrased in the following form:
If we push the surface of (our model of) an elastic body with a constant force at a single point, and look for the deformation this force produces, the indentation increases logarithmically in time due to the relaxation processes considered by the model. 
In this respect, I notice that the logarithmic increase is crucially dependent on the dimensionality of the surface. For instance, in the case of a line (i.e., the border of a half plane) the same relaxation mechanism would produce a displacement that grows with time as $t^{1/4}$ (this dependence appearing when summing up the $q$ modes of Eq. (\ref{uq}) in a one-dimensional geometry), which is (at least in principle) discernible from a logarithmic increase. It thus remains to be seen if there are experimental realizations on this confined geometry configuration that allows to test this prediction. 

I now want to discuss in more detail the analytical expression in Eq. (\ref{ndet2}), that we saw is well verified in the simulations, and compare it with available experimental results. 
One of the most detailed experimental studies of temporal effects in friction measurement have been provided by Baumberger and co-workers in a series of papers \cite{caroli,berthoud,baum1,baum2}.
One important parameter they consider, is the coefficient defined as the derivative of the static friction coefficient $\mu_s$ with respect to the logarithm of the hold time\cite{caroli,marone0}, namely $B\equiv \frac{d \mu_s}{d\ln(t)}$. This hold time is the time during which the two surfaces are left in rest contact, before the friction force necessary to start sliding is measured.
 This coefficient (that is independent of the time unit chosen) is typically found to have a conserved value for a variety of materials, in a range close to 
$10^{-2}$.

Following Tabor\cite{bt}, the friction force $F_{fr}$ between two solids can be written as $F_{fr}=\sigma_S A$, where $\sigma_S$ is the so-called shear strength of the interface and $A$ the real area of contact. Using also the standard expression $F_{fr}=\mu_s F_N$, we can write $\mu_s=A\sigma_S/F_N$, where we see that the friction coefficient is directly related to the contact area. In particular, if we assume that $\sigma_S/F_N$ takes some constant value, we can write
\begin{equation}
\frac{1}{\mu_s}\frac{d\mu_s}{d\ln t}=\frac{1}{A}\frac{dA}{d\ln t}
\end{equation}
and since $\mu_s$ is typically of order one, we can roughly write
\begin{equation}
B\simeq \frac{1}{A}\frac{dA}{d\ln t}\simeq \frac{d(A/A(t=0))}{d\ln t}
\end{equation}
This form gives a direct access to the $B$ coefficient as predicted by the model presented in this paper.
We see first of all (from Eq. \ref{ndet2}) that $B$ does not depend on the relaxation coefficient $R$. This seems a bit surprising but is not contradictory: although the contact area increase in time is produced by a non-zero value of $R$, this coefficient enters directly in setting the time scale (as it is obvious from Eq. (\ref{u0s})) but the logarithmic derivative of contact area is independent of it. Once this fact is recognized, we may ask to what extent the value of $B$ in our model can be considered to be constant, independent of other parameters, and if this is the case, if this value is in the experimentally observed range $B\sim 10^{-2}$. In this respect, it seems that the answer is negative, since the coefficient is directly related to the ratio $k_1/k_0$ (Eq. (\ref{ndet2})), which can in principle be set arbitrarily. However, I already stressed the fact that the description of our elastic surface is not very accurate. In fact, the elastic properties of the surface of an elastic body, that I consider isotropic for simplicity, can be characterized by the values of one elastic constant (the Young modulus $Y$, for instance) and are only weakly
dependent of a second parameter, namely the Poisson ratio $\nu$. 
The dependence of $B$ on the dimensionless ratio $k_1/k_0$ is an indication that in a hypothetical more accurate description of the elastic body, $B$ cannot depend explicitly on the value of $Y$. Only a dependence on $\nu$ can exist.
Then I expect in fact a rather weak dependence of $B$ on the elastic parameters of the body, and then a conserved value of $B$ for different materials. Whether this conserved value is compatible with the experimental value $\sim 10^{-2}$ has to be answered once a more realistic description of the elastic surface of the body is done.

Going a step further we must discuss the effect of temperature. The Brechet-Estrin analysis \cite{brechet} predicts that the $B$ coefficient must be temperature dependent, with values that increase with increasing temperature.
Berthoud {\it et al.}\cite{berthoud} were able to observe systematic variations of this coefficient with temperature in different experimental situations, that are roughly compatible with the predictions of Brechet and Estrin. Typically, an increase of $B$ (up to a factor of roughly ten) was observed when the glass temperature of the material was approached.
In the present model it is  not obvious where a temperature dependence can enter that alter the value of the $B$ coefficient. The only obvious temperature dependent parameter is the relaxation coefficient $R$, but we have already
seen that $B$ is independent of $R$.

\begin{figure}
\centerline{\includegraphics[width=.5\textwidth]{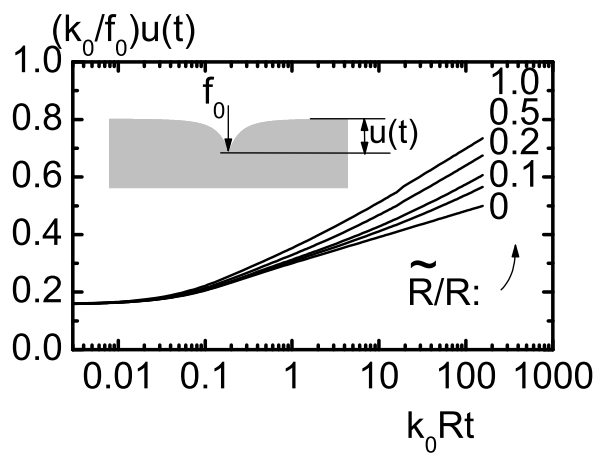}}
\caption{Depth of an indentation experiment on the model, under a constant load $f_0$, as a function of time, for $k_1/k_0=3$, in the presence of creep relaxation and viscoelastic relaxation, measured respectively by the coefficients $R$ and $\tilde R$. (see Eq.  (\ref{dosrs})). Note the increase in the slope of the asymptotic logarithmic behavior, as $\tilde R/R$ is increased.
}
\label{indent}
\end{figure}

A possible way out to this situation may involve to consider the possibility of other mechanisms of relaxation, in addition to the one considered in Eq. (\ref{u0s}). For instance, in addition to the present mechanism that responds to the fluctuations of the forces on the $k_1$ springs, we can add a term that is directly dependent of the force itself.
This would generalize Eq. (\ref{u0s}) to an equation of the type

\begin{equation}
\frac{du^0({\bf r})}{dt}=R\nabla^2\frac{\delta E}{\delta u^0({\bf r})}-\tilde R\frac{\delta E}{\delta u^0({\bf r})}
\label{dosrs}
\end{equation}

One of the main qualitative difference caused by the inclusion of the last term is that under the action of a constant 
value of $\overline {u^0}$, the value of $F_N$ goes to zero at very large times (contrary to the finite value reached in the presence of the first term alone). In this sense the last term represents a ``viscous" relaxation in the system, that may have a progressively larger effect as the temperature increases towards the glass temperature of the system.
This suggests that $\tilde R/R$ may be considered to be an increasing function of temperature.

The effect of this term can be seen in Fig. \ref{indent}. There I show the time dependence of the indentation depth for a point contact on the surface of the elastic body for increasing values of $\tilde R/R$. These results are obtained by solving analytically the system equations for $u_q$ as I did in Section V, using Eq. (\ref{dosrs}) instead of (\ref{u0s}), and Fourier inverting numerically the result. We see that the logarithmic increase of $u$ in time is conserved in the presence of the $\tilde R$ term, and the slope increases as $\tilde R$ increases. This slope measures directly the value of $B$ in the model. It can be easily shown that $B$ doubles its value as $\tilde R/R$ goes from zero to large values. This means that the consideration of alternative relaxation mechanisms can justify a variation of the $B$ coefficient with temperature. Whether the change of the $B$ coefficient observed by Berthoud {\em et al.}\cite{berthoud} 
is related to a change of relaxation mechanism or not remains to be investigated 
further, both theoretically and experimentally.



This research was financially supported by Consejo Nacional de Investigaciones Cient\'{\i}ficas y T\'ecnicas (CONICET), Argentina. Partial support from PICT 32859/2005 (ANPCyT, Argentina) is also acknowledged.\\


\begin{thebibliography}{4}
\bibitem{notita}This condition in particular allows a numerical determination of the fully relaxed contact area by
imposing a constant force onto the $k_1$ springs, and relaxing towards mechanical stability. This is the way in which
the ``fully relaxed" results in Figs. 2, 4, and 7 (last panel) have been obtained.
\end{thebibliography}
\end{document}